\documentclass[runningheads,envcountsame]{llncs}

\usepackage[pdftex]{graphicx,color}
\usepackage{latexsym,amsmath,amssymb,amstext,bussproofs}

\newcommand{\Leadsto}{\mathrm{leadsto}}
\newcommand{\Conv}{\mathrm{conv}}
\newcommand{\Div}{\mathrm{div}}
\newcommand{\Dlf}{\mathrm{dlf}}

\title{A Complete Fragment of LTL(EB)}

\author{Flavio Ferrarotti\inst{1}\thanks{The research of Flavio Ferrarotti has been funded by the Federal Ministry for Climate Action, Environment, Energy, Mobility, Innovation and Technology (BMK), the Federal Ministry for Digital and Economic Affairs (BMDW), and the State of Upper Austria in the frame of the COMET Module Dependable Production Environments with Software Security (DEPS) within the COMET - Competence Centers for Excellent Technologies Programme managed by Austrian Research Promotion Agency FFG.} \and Peter Rivi\`{e}re\inst{2} \and Klaus-Dieter Schewe\inst{2} \and Neeraj Kumar Singh\inst{2} \and Yamine A\"{\i}t Ameur\inst{2}}
\authorrunning{F. Ferrarotti, K.-D. Schewe}
\institute{Software Competence Centre Hagenberg, Austria,
\email{flavio.ferrarotti@scch.at} \and
Institut Nationale Polytechnique de Toulouse / ENSEEIHT, Toulouse, France, \email{\{peter.riviere$\mid$nsingh$\mid$yamine\}@enseeiht.fr}, \email{kd.schewe@liwest.at}}

\begin{document}

\maketitle 

\begin{abstract}

The verification of liveness conditions is an important aspect of state-based rigorous methods. This article investigates this problem in a fragment $\square$LTL of the logic LTL(EB), the integration of the UNTIL-fragment of Pnueli's linear time temporal logic (LTL) and the logic of Event-B, in which the most commonly used liveness conditions can be expressed. For this fragment a sound set of derivation rules is developed, which is also complete under mild restrictions for Event-B machines.


\end{abstract}

\section{Introduction}

State-based consistency conditions such as state or transition invariants are included in rigorous state-based methods such as B  \cite{abrial:2005}, Abstract State Machines (ASMs) \cite{boerger:2003}, TLA$^+$ \cite{lamport:2002} or Event-B \cite{abrial:2010}, just to mention the most important ones. The verification of such conditions is well supported by appropriate logics such as the logics for Event-B  \cite{schmalz:eth2012}, TLA$^+$ \cite{merz:cai2003}, deterministic ASMs \cite{staerk:jucs2001} and arbitrary non-deterministic ASMs \cite{ferrarotti:amai2018}. As the logics can be defined as definitional extensions of first-order logic with types, they are complete \cite{vaananen:sep2019}, which is a valuable asset for verification.

However, besides such consistency conditions other liveness conditions are likewise important. These conditions comprise among others (conditional) {\em progress}, i.e. any state satisfying a condition $\varphi$ is always followed eventually (i.e. some time later) by a state satisfying $\psi$ (for $\varphi = \textbf{true}$ we obtain unconditional progress, sometimes also called {\em existence}), or {\em persistence}, i.e. eventually a condition $\varphi$ will hold forever. Their verification requires the reasoning about complete runs or traces of a specification, i.e. they are intrinsically connected to temporal logic \cite{kroeger:2008}. Specific temporal logics that have been used already for a long time for the verification of desirable liveness properties of sequential and concurrent (interleaved) systems are Linear-Time Temporal Logic (LTL) \cite{pnueli:focs1977} and Computation Tree Logic (CTL) \cite{clarke:lop1981}. For many liveness conditions it suffices to consider only the UNTIL-fragment of LTL \cite{manna:scp1984}. These logics, however, have been defined as temporal propositional logics, hence require extensions when used in connection with any of the rigorous methods mentioned above.

Hoang and Abrial started to investigate the use of the UNTIL-fragment of LTL in connection with Event-B \cite{hoang:icfem2011}, and recently Rivi\`{e}re et al. provided a tool support using the EB4EB extension of RODIN \cite{riviere:nfm2023}. Defining a logic that integrates LTL with the logic of Event-B is rather straightforward; instead of propositions consider formulae that can be defined for a given Event-B machine. In doing so, NEXT-formulae are already covered by the logic of Event-B, and hence the omission of the NEXT modal operator does no harm. Note that a similar argument is used for the logic of TLA$^+$ \cite{merz:cai2003}.

Naturally, with this integration we obtain a first-order modal logic, which cannot be complete. Nonetheless, Hoang and Abrial discovered a few derivation rules for invariance, existence, progress and persistence and proved their soundness, thus showing that if certain variant terms can be derived in an Event-B machine, such liveness properties can be verified. However, the question remains, if the derivation rules are also complete, at least for a well-defined fragment of LTL(EB).

Here we approach this problem defining a fragment $\square$LTL of LTL(EB), for which we use the type of formulae in existence, progress and persistence conditions as defining constructors. Then we redo the soundness proofs from \cite{hoang:icfem2011} with slight generalisations. We add a few standard derivation rules for modal formulae, and show that the set of all these derivation rules is also complete for the $\square$LTL fragment. To be precise, all valid formulae are derivable, if appropriate variant terms can be defined in Event-B machines, a problem that was left open in \cite{hoang:icfem2011}, and the machines are tail-homogeneous, which means that the intrinsic non-determinism in Event-B is restricted. The tail-homogeneity restriction is necessary, as LTL is a linear time temporal logic and as such does not work well together with branching traces.

In order to show that such variant terms, which are needed for proofs of convergence or divergence properties, always exist, we use a streamlined version of Event-B, in which all sets used in an Event-B machine are defined as subsets of a single set $\text{HF}(A)$ of hereditarily finite sets over a finite set $A$ of atoms. In this case all values to which state variables can be bound, become hereditarily finite sets. This is similar to the special ASMs used in Choiceless Polynomial Time \cite{blass:apal1999}, and it is well known that this is no loss of generality, as sets can always be encoded in this way \cite{barwise:1975}. The advantage is that a few operators on $\text{HF}(A)$ suffice to express all terms and consequently also all state formulae, which is exploited is the proof of the existence of variant terms in conservatively refined machines. That is, the logic $\square$LTL is sound and complete for proofs on sufficiently refined, tail-homogeneous Event-B machines.

\paragraph{Organisation of the Paper.}

In Section \ref{sec:ltleb} we first give a brief introduction to Event-B and to LTL, which give rise to the definition of the logic LTL(EB). As in \cite{hoang:icfem2011} we highlight specific formulae for state transitions, deadlock-freeness, convergence and divergence. The former two of these are merely definitional extensions of first-order logic. For convergence and divergence, however, we need variant terms, so we will show that such terms always exist for sufficiently refined machines. For reason of space limitations full proofs have been outsourced into an appendix. We continue in Section \ref{sec:fragment} with the definition of the $\square$LTL fragment, the presentation of a set of derivation rules and the proofs of soundness and completeness for tail-homogeneous machines. The most common liveness properties can be expressed in the fragment. We conclude with a brief discussion of the value and limitations of the results in Section \ref{sec:schluss}.

\section{The Logic LTL(EB)}\label{sec:ltleb}

The logic LTL(EB) integrates the logic of Event-B into the UNTIL-fragment of LTL, which results in a first-order temporal logic. LTL(EB) formulae are interpreted over sequences of states, and thus can be used for stating and proving dynamic properties of computations of Event-B machines. We assume familiarity with the well-known Event-B method \cite{abrial:2010}, and only recall some basic concepts fixing the notation needed in the remainder of this paper.

\subsection{Event-B}

The basic schema views an Event-B \emph{machine} (a.k.a. program or model) as offering a set of \emph{events} $E$ (operations), which are executed one at a time, i.e. only one event can be fired at any time point. Events are usually parameterised, which results in invariant-preserving state transitions. The firing of events is the only means for updating states. 

A \emph{state} in Event-B is formed by a finite tuple of variables $\bar{v}$ taking values in certain sets with (explicitly or implicitly defined) auxiliary functions and predicates. The sets are usually specified in an Event-B \emph{context}, but syntactical details are of no importance here. 

For the sake of streamlining our proofs, however, we will adopt that all these sets are subsets of the set $\text{HF}(A)$ of {\em hereditarily finite sets} over a finite set $A$ of atoms. That is, if $\mathcal{P}$ denotes the powerset operator and we define inductively $\mathcal{P}^0(A) = A$ and $\mathcal{P}^{i+1}(A) = \mathcal{P}(\bigcup_{j \le i} \mathcal{P}^j(A))$, then we have
\[ \text{HF\/}(A) = \bigcup_{i < \omega} \mathcal{P}^i(A) = A \cup \mathcal{P}(A) \cup \mathcal{P}(A \cup \mathcal{P}(A)) \cup \dots . \]
In particular, all values to which state variables in $\bar{v}$ can be bound are then hereditarily finte sets. It is well known that this is no loss of generality, as sets used in Event-B contexts can be encoded this way \cite{barwise:1975}. In particular, $\text{HF}(A)$ contains a standard model of the set of natural numbers.

Then we use set operators $\in$, $\emptyset$, \textit{Atoms\/}, $\bigcup$, \textit{TheUnique\/} and \textit{Pair\/}. The predicate $\in$ and the constant symbol $\emptyset$ are interpreted in the obvious way, and \textit{Atoms\/} is interpreted by the set $A$ of atoms. If $a_1 ,\dots, a_k$ are atoms and $b_1 ,\dots, b_\ell$ are sets, then $\bigcup \{ a_1 ,\dots, a_k, b_1 ,\dots, b_\ell \} = b_1 \cup\dots\cup b_\ell$. For $b = \{ a \}$ we have $\textit{TheUnique\/}(b) = a$, otherwise $\textit{TheUnique\/}(b) = \emptyset$ denoting that it is undefined. Furthermore, we have $\textit{Pair\/}(a,b) = \{ a, b \}$. As shown in \cite[Lemma~13]{blass:apal1999} the usual set operations (union, intersection, difference) can be expressed by this term language.

Each \emph{event} $e_i \in E$ has the form $\mathbf{any} \, \bar{x} \, \mathbf{where} \, G_i(\bar{x}, \bar{v}) \, \mathbf{then} \, A_i(\bar{x}, \bar{v}, \bar{v}') \, \mathbf{end}$, where $\bar{x}$ are the \emph{parameters}, the first-order formula $G_i(\bar{x}, \bar{v})$ is the \emph{guard} and $A_i(\bar{x}, \bar{v}, \bar{v}')$ is the \emph{action} of $e_i$.

An event $e_i$ is \emph{enabled} in state $S$ with variables $\bar{v}$, if there are values for its parameters $\bar{x}$ that make its guard $G_i(\bar{x}, \bar{v})$ hold in $S$.  Otherwise, the
event is \emph{disabled}. If all events of a machine $M$ are disabled in an state $S$, then $M$ is said to be \emph{deadlocked} in $S$. 

An Event-B {\em action} $A_i(\bar{x}, \bar{v}, \bar{v}')$ is a list of assignments of the form $v :\!| \, P(\bar{x}, \bar{v}, v')$, where $v$ is a variable that occurs in the tuple $\bar{v}$ of state variables and $P$ is a (so called) \emph{before-after predicate} relating the value of $v$ (before the action) and $v'$ (after the action). The value assigned to $v$ is chosen non-deterministically\footnote{It is commonly assumed that this choice is external, i.e. the values are provided by the environment and not by the machine itself.} from the set of values  $v'$ that satisfy $P(\bar{x}, \bar{v}, v')$. Thus, each action $A_i(\bar{x}, \bar{v}, v')$ corresponds to a before-after predicate $P_{A_i}(\bar{x}, \bar{v}, \bar{v}')$ formed by the conjunction of all before-after predicates of the assignments in $A_i(x, \bar{v}, v')$. These predicates are defined using the language of set theory, i.e. first order logic with the binary relation-symbol $\in$. There are several dedicated notations for (typed) set operations such as set comprehension, cartesian product, etc. We do not discuss them here as they are definable in first-order logic and well known.   

An action also allows assignments of the form $v := E(\bar{x},\bar{v})$ and $v :\in E(\bar{x},\bar{v})$, where $E(\bar{x},\bar{v})$ is an expression. The assignment $v := E(\bar{x},\bar{v})$ deterministically assigns the value of $E(\bar{x},\bar{v})$ to $v$, while $v :\in E(\bar{x},\bar{v})$  non-deterministically assigns an element of the set $E(\bar{x},\bar{v})$ to $v$. However, these two assignments are merely shortcuts of assignments in the general form $v :\!| \, P(\bar{x}, \bar{v}, v')$ with before-after predicates $v' = E(\bar{x},\bar{v})$ and $v' \in E(\bar{x},\bar{v})$, respectively.  

The values of the variables in the initial state are set by means of an special event called  $\mathit{init}$ that has no parameters or guard. The values that variables $\bar{v}$ can take in a state are constrained by invariants $I(\bar{v})$. These invariants are to hold in every reachable state, which is achieved by proving that they are \emph{established} by the initialisation event $\mathit{init}$ and subsequently \emph{preserved} by all other events.  

Given an event $e$, we say that a state $S^\prime$ is an $e$-successor state of $S$, if $S^\prime$ is a possible
after-state of the firing of $e$ from the before-state $S$. Lifting the definition to a machine $M$, we say that $S^\prime$ is an $M$-successor state of $S$ if there exists an event $e$ of $M$ such that $S^\prime$ is an $e$-successor of $S$.

Event-B defines a \emph{computation} $\tau$ (or \emph{trace} in Event-B terminology) of a machine $M$ as a (possible infinite) sequence of states $S_0, S_1, S_2, \ldots$ such that $S_0$ is an initial state satisfying the after predicate defined by the event $\mathit{init}$, and for every pair of consecutive states $S_i$, $S_{i+1}$ in $\tau$, there is an event $e$ such that $S_{i+1}$ is an $e$-successor state of $S_i$. If the computation $\tau$ ends in an state $S_n$, i.e. the computation terminates, then $M$ must be deadlocked in $S_n$.

Note that in general for a machine $M$ and an initial state $S_0$ there exist multiple traces. This includes the possibility that infinite and finite traces occur together. As this is inconvenient for the proof of liveness properties, we introduce an additional event \textit{ext}, by means of which all traces become infinite. The guard of \textit{ext\/} is $\bigwedge\limits_{e_i \in E} \forall \bar{x} \neg G_i(\bar{x}, \bar{v})$, so the event \textit{ext\/} is enabled iff no other event $e_i \in E$ is enabled. The action of \textit{ext\/} is given by a list of assignments $v :| v^\prime = v$ for all variables $v$ in $\bar{v}$, so a firing of \textit{ext\/} does not change the state. That is, all infinite traces remain unchanged, and all finite traces are infinitely extended by repeating the final state. We call a machine $\tilde{M}$ that extends $M$ by such an event \textit{ext\/} the {\em trivial extension} of $M$.

Given a machine $M$ we immediately obtain a language $\mathcal{L}$ of {\em state formulae} of $M$. If the state variables of $M$ are $\bar{v}$, then the set of basic well formed formulae of ${\cal L}$ is formed by the set of first-order logic formulae over some signature $\tau$ with all free variables among those in $\bar{v}$. Such formulae are interpreted in any state $S$ of $M$, where the value of the free (state) variables is interpreted by the value of these variables in $S$.

For any state formula $\varphi$ we can also define the first-order formula
\[ \mathcal{N}\varphi(\bar{v}) \equiv \bigwedge_{e_i \in E} \forall \bar{x} \Big ( G_i(\bar{x}, \bar{v}) \rightarrow \forall \bar{v}^\prime ( P_{A_i}(\bar{x}, \bar{v}, \bar{v}^\prime) \rightarrow \varphi(\bar{v}^\prime) ) \Big ) \; . \]

The formula $\mathcal{N}\varphi$ holds in a state $S$, if the formula $\varphi$ holds in all $M$-successor states of $S$.

\subsection{Linear Time Temporal Logic}

Linear time temporal logic (LTL) is a well known propositional temporal logic with modal operators referring to time~\cite{pnueli:focs1977}. This logic has been used for the formal verification of computer programs for a long time \cite{manna:scp1984}; only a fragment suffices to express the most important liveness conditions. Thus formulae are built from propositions, the usual Boolean operators $\neg, \wedge, \vee, \rightarrow$, and modal operators. There are different ways to define the basic modal operators of LTL; we will use only the three constructors $\square$ (always), $\lozenge$ (eventually, i.e. sometimes in the future), and $\mathcal{U}$ (until) \cite{manna:scp1984}. 

Here we use state formulae of an Event-B machine instead of propositions, which defines the logic LTL(EB). Then the set of {\em well-formed formula} of LTL(EB) is defined as the closure of the set $\cal L$ of basic state formulae under the usual Boolean operators $\neg, \wedge, \vee, \rightarrow$ and modal operators $\square$ (always), $\lozenge$ (eventually) and ${\cal U}$ (until). Note that $\lozenge \, \varphi$ is merely a shortcut for $\mathbf{true} \, \mathcal{U} \, \varphi$, and $\square \, \varphi$ is a shortcut for $\neg \lozenge \neg \varphi$, as can be easily proven from the semantics we define below. As we saw above, the absence of the LTL modal operator $\mathcal{N}$ (next) does not matter, as conditions that are to hold in successor states are already covered by the first-order logic of Event-B.

LTL(EB) formulae are interpreted over traces of Event-B machines, which we now define formally. Let $\sigma$ be a non-empty sequence of states $S_0, S_1, \ldots$. Let $\ell(\sigma) = n$ if the sequence $\sigma$ is finite and has length $n$, and $\ell(\sigma) = \omega$ if it is infinite. Let $\sigma^{(k)}$ for $0 \leq k \leq \ell(\sigma)$ denote the sequence $S_k, S_{k+1}, \ldots$ obtained from $\sigma$ by removing its first $k$ elements. An LTL(EB) formula $\varphi$ is satisfied by a sequence of states $\sigma$ with state variables $\bar{v}$, denoted $\sigma \models \varphi$, iff the following holds:

\begin{itemize}

    \item If $\varphi \in {\cal L}$, i.e. if $\varphi$ is a first-order formula with all its free variables in $\bar{v}$, then $\sigma \models \varphi$ iff the values assigned by the first state $S_0$ of $\sigma$ to the variables $\bar{v}$ satisfies $\varphi$.
    
    \item If $\varphi$ is of the form $\neg \psi$, $\psi \wedge \chi$, $\psi \vee \chi$ or $\psi \rightarrow \chi$, where $\psi$ and $\chi$ are well-formed LTL(EB) formulae, then $\sigma \models \varphi$ iff $\sigma \not\models \psi$, $\sigma \models \psi$ and $\sigma \models \chi$, $\sigma \models \psi$ or $\sigma \models \chi$, or $\sigma \models \chi$ whenever $\sigma \models \psi$, respectively.  
    
    \item If $\varphi$ is of the form $\square \, \psi$, then $\sigma \models \varphi$ iff for all $0 \leq k \leq \ell(\sigma)$ it holds that $\sigma^{(k)} \models \psi$. That is, $\sigma \models \square \, \psi$ iff all states of $\sigma$ satisfy $\psi$.
    
    \item If $\varphi$ is of the form $\lozenge \, \psi$, then $\sigma \models \varphi$ iff there is a $0 \leq k \leq \ell(\sigma)$ such that $\sigma^{(k)} \models \psi$. That is, $\sigma \models \lozenge \, \psi$ iff some state of $\sigma$ satisfies $\psi$.
    
    \item If $\varphi$ is of the form $\psi \, {\cal U} \, \chi$, then $\sigma \models \varphi$ iff there is a $0 \leq k \leq \ell(\sigma)$ such that $\sigma^{(k)} \models \chi$ and for all $0 \leq i < k$ it holds that $\sigma^{(i)} \models \psi$. That is, $\sigma \models \psi \, {\cal U} \, \chi$ iff some state $S_k$ satisfies $\chi$ and all states in the sequence $\sigma$ until $S_k$ (where $S_k$ is not included) satisfy $\psi$\footnote{Note that this is a strong interpretation of until-formulae, as it includes that $\chi$ eventually holds. If this requirement is dropped, i.e. we allow $\psi$ to hold forever, we obtain a weak notion of until-formulae, usually denoted as $\psi \, {\cal W} \, \chi$.}.
    
\end{itemize}

For an Event-B machine $M$ with state variables $\bar{v}$ let $\varphi$ be an LTL(EB) formulae such that a variable $v_i$ is free in $\varphi$ iff $v_i$ appears in $\bar{v}$. We say that $M$ \emph{satisfies} property $\varphi$ (denoted as $M \models \varphi$) iff all possible traces of $M$ satisfy $\varphi$.

\subsection{Variant Formulae}

Our aim is to develop a proof system for LTL(EB) based on the derivation rules in \cite{hoang:icfem2011} for reasoning about liveness properties in Event-B. The models in this proof system correspond to sequences of Event-B states. Thus, a sequence of Event-B states $\sigma$ is a {\emph model} of an LTL(EB) formula $\varphi$ iff $\sigma \models \varphi$ holds. If $\Psi$ is a set of formulae, then $\sigma$ models $\Psi$ (denoted as $\sigma \models \Psi$) iff $\sigma \models \varphi$ holds for every $\varphi \in \Psi$. A formula $\varphi$ is a \emph{logical consequence} of a set of formulae $\Psi$ (denoted as $\Psi \models \varphi$) iff for all sequences of Event-B states $\sigma$ whenever $\sigma \models \Psi$holds, then also $\sigma \models \varphi$. 

Let $\mathfrak{R}$ be a set of axioms and inference rules. A formula $\varphi$ is \emph{derivable} from a set $\Psi$ of formulae using $\mathfrak{R}$ (denoted as $\Psi \vdash_{\mathfrak{R}} \varphi$, or simply $\Psi \vdash \varphi$ when $\mathfrak{R}$ is clear from the context) iff there is a deduction from formulae in $\Psi$ to $\varphi$ that only applies axioms and inference rules from $\mathfrak{R}$.

The notion of {\em validity} applies to a given Event-B machine $M$ as explained in this section. That is, a formula $\varphi$ is \emph{valid for} $M$ (denoted as $M \models \varphi$) iff $\sigma \models \varphi$ holds for all computations $\sigma$ of $M$. If $M \models \varphi$ can be derived from a set of axioms and inference rules $\mathfrak{R}$, then we write $M \vdash_{\mathfrak{R}} \varphi$.

Before we recall and extend the LTL(EB) proof rules from~\cite{hoang:icfem2011}, we define certain common LTL(EB) formulae used in those rules. For this let $M$ be an Event-B machine with a set $E$ of events and state variables $\bar{v}$. For each $e_i \in E$, let $G_i(\bar{x}, \bar{v})$ and $P_{A_i}(\bar{x}, \bar{v}, \bar{v}')$ denote the guard and the before-after predicate corresponding to the action $A_i(\bar{x}, \bar{v}, \bar{v}')$ of event $e_i$, respectively. 

For state formulae $\varphi_1$ and $\varphi_2$ we say that \emph{$M$ leads from $\varphi_1$ to $\varphi_2$} iff the following sentence holds:
\[ \Leadsto(\varphi_1, \varphi_2) \equiv  \bigwedge_{e_i \in E} \forall \bar{v} \bar{x} \bar{v}'  \Big( \varphi_1(\bar{v}) \wedge G_i(\bar{x}, \bar{v}) \wedge P_{A_i}(\bar{x}, \bar{v}, \bar{v}') \rightarrow \varphi_2(\bar{v}') \Big).\]

It implies that for every computation $\sigma$ of $M$ and every $0 \leq k \leq \ell(\sigma)$, it holds that whenever $\sigma^{(k)} \models \varphi_1$ then $\sigma^{(k+1)} \models \varphi_2$. We therefore call such formulae {\em state transition formulae}.

For a state formula $\varphi$ we say that \emph{$M$ is deadlock-free in $\varphi$} iff following sentence is satisfied:
\[ \Dlf(\varphi) \equiv \forall \bar{v} \Big( \varphi(\bar{v}) \rightarrow \exists \bar{x} \big(\bigvee_{e_i \in E} G_i(\bar{x}, \bar{v}) \big) \Big). \]

It expresses that whenever a state satisfies $\varphi$, there is at least one event $e_i$ of $M$ that is enabled. In particular, $\sigma^{\ell(\sigma)} \models \neg \varphi$ holds for every finite computation $\sigma$ of $M$. 

State transition formulae $\Leadsto(\varphi_1, \varphi_2)$ and deadlock-freeness formulae $\Dlf(\varphi)$ express properties of traces of an Event-B machine, but nonetheless they are defined in first-order logic. As first-order logic is complete, such formulae are derivable using a standard set of derivation rules for first-order logic. This also holds for the logic of Event-B, which is first-order with types. 

This will no longer be the case for convergence and divergence formulae $\Conv(\varphi)$ and $\Div(\varphi)$, respectively, which we define next.

For a state formula $\varphi$ we say that \emph{$M$ is convergent in $\varphi$} iff for every computation $\sigma$ of $M$ there is no $k$ such that $\sigma^{(\ell)} \models \varphi$ holds for all $\ell \ge k$. Note that if a machine $M$ satisfies $\Conv(\varphi)$, then its trivial extension $\tilde{M}$ also satisfies $\Conv(\varphi)$ and vice versa. Hoang and Abrial showed that $\Conv(\varphi)$ can be derived by means of {\em variant} terms \cite[p.~460]{hoang:icfem2011}. As we adopted that all values are hereditarily finite sets in $\text{HF}(A)$, we can exploit a canonical partial order $\le$ on $\text{HF}(A)$, defined by $x \le y$ iff $\text{TC}(x) \subseteq \text{TC}(y)$, where $\text{TC}(x)$ denotes the transitive closure of $x$. This order is well-founded with $\emptyset$ as the smallest element, and it subsumes set inclusion $\subseteq$ as well as the reflexive, transitive closure of the membership relation $\in$. 

Suppose that $t(\bar{v})$ is a well-formed term called {\em variant} and $\text{var}_c(t,\varphi)$ denotes the formula
\[ \bigwedge_{e_i \in E} \forall \bar{v} \bar{x} \bar{v}' \Big( \varphi(\bar{v}) \wedge G_i(\bar{x}, \bar{v}) \rightarrow \big( t(\bar{v}) \neq \emptyset \wedge ( P_{A_i}(\bar{x}, \bar{v}, \bar{v}') \rightarrow t(\bar{v}') < t(\bar{v}) ) \big) \Big) .\]

That is, a state $S$ satisfies $\text{var}_c(t,\varphi)$ iff whenever $\varphi$ holds and an event $e_i$ is enabled, we have that $t(\bar{v})$ evaluates to a non-empty set, and an execution of $e_i$ decreases $t(\bar{v})$. Hoang and Abrial showed the following:

\begin{lemma}

If $\text{var}_c(t,\varphi)$ is a valid formula and no trace of $M$ terminates in a state satisfying $\varphi$, then $M$ is convergent in $\varphi$, in other words, the derivation rule\\
\begin{minipage}{12.5cm}
\begin{prooftree}
\AxiomC{$M \vdash \text{var}_c(t,\varphi)$}
\AxiomC{\hspace*{-5mm}$M \vdash \, \Dlf(\varphi)$}
\LeftLabel{{\rm CONV}: \quad}
\BinaryInfC{$M \vdash \Conv(\varphi)$}
\end{prooftree}
\end{minipage}
is sound.

\end{lemma}

\begin{proof}

If $M \models \Dlf(\varphi)$ holds, then every finite trace of $M$ terminates in a state satisfying $\neg\varphi$.
If $M \models \text{var}_c(t,\varphi)$, but $M \not\models \Conv(\varphi)$, then by the definitions of $\Conv(\varphi)$ and $\text{var}_c(t,\varphi)$ there is an infinite subsequence $S_i, S_{i+1}, \dots$ of states of some trace $\sigma$ of $M$ such that for the values $b_j$ of $t(\bar{v})$ in $S_j$ we have $b_i > b_{i+1} > \dots$. As the order $\le$ on $\text{HF}(A)$ is well-founded, this is not possible.\qed

\end{proof}

Analogously, for a state formula $\varphi$ we say that \emph{$M$ is divergent in $\varphi$} iff for every infinite computation $\sigma$ of $M$ there is a $k$ such that $\sigma^{(k')} \models \varphi$ for every $k' \geq k$. Note that if $\tilde{M} \models \Div(\varphi)$ holds for the trivial extension, then also $M \models \Div(\varphi)$ holds, but the converse is false in general. However, if $M \models \Div(\varphi)$ holds and all finite traces terminate in a state satisfying $\varphi$, then also $\tilde{M} \models \Div(\varphi)$ follows.

Hoang and Abrial also showed that $\Div(\varphi)$ can be derived by means of {\em variant} terms \cite[p.~460]{hoang:icfem2011}. For this suppose that $t(\bar{v})$ is a well-formed term called {\em variant} and let $\text{var}_d(t,\varphi)$ denote the formula 
\begin{gather*}
\bigwedge_{e_i \in E} \forall \bar{v} \bar{x} \bar{v}' \Big( \neg \varphi(\bar{v}) \wedge G_i(\bar{x}, \bar{v}) \rightarrow \big( t(\bar{v}) \neq \emptyset \wedge ( P_{A_i}(\bar{x}, \bar{v}, \bar{v}') \rightarrow t(\bar{v}') < t(\bar{v}) ) \big) \Big) \\
\wedge \Big( \varphi(\bar{v}) \wedge G_i(\bar{x}, \bar{v}) \wedge P_{A_i}(\bar{x}, \bar{v}, \bar{v}') \rightarrow t(\bar{v}') \le t(\bar{v}) ) \Big) .
\end{gather*}

That is, a state $S$ satisfies $\text{var}_d(t,\varphi)$ iff whenever $\neg\varphi$ holds and an event $e_i$ is enabled, we have that $t(\bar{v})$ evaluates to a non-empty set, and an execution of $e_i$ decreases $t(\bar{v})$, and whenever $\varphi$ holds and an event $e_i$ is enabled, we have that an execution of $e_i$ does not increase $t(\bar{v})$.

\begin{lemma}

If $\text{var}_d(t,\varphi)$ is a valid formula, then $M$ is divergent in $\varphi$, in other words, the derivation rule\\
\begin{minipage}{12.5cm}
\begin{prooftree}
\AxiomC{$M \vdash \text{var}_d(t,\varphi)$}
\LeftLabel{{\rm DIV}: \quad}
\UnaryInfC{$M \vdash \Div(\varphi)$}
\end{prooftree}
\end{minipage}
is sound.

\end{lemma}

\begin{proof}

If $M \models \text{var}_d(t,\varphi)$, but $M \not\models \Div(\varphi)$, then by the definition of $\Div(\varphi)$ there is an infinite trace $\sigma = S_0, S_1, \dots$ of $M$ such that for all $i$ there is some $k_i \ge i$ with $S_{k_i} \models \neg\varphi$. Let $b_j$ be the value of $t(\bar{v})$ in $S_j$. Then by the definition of $\text{var}_d(t,\varphi)$ we have $b_i \ge b_{i+1}$ and $b_{k_i} > b_\ell$ for all $i$ and all $\ell > k_i$. As the order $\le$ on $\text{HF}(A)$ is well-founded, this is not possible.\qed

\end{proof}

We call the first-order formulae $\text{var}_c(t,\varphi)$ and $\text{var}_d(t,\varphi)$ {\em variant formulae}. Then the derivation rules CONV and DIV allow us to reduce proofs of convergence and divergence properties to proofs of formulae in first-order logic, if we can find appropriate variant terms. We will show that this is always possible. Let $\text{var}_c(M, t,\varphi)$ and $\text{var}_d(M, t,\varphi)$ denote the formulae obtained from the sentences $\text{var}_c(t,\varphi)$ and $\text{var}_d(t,\varphi)$, respectively, by making $\bar{v}$ and $\bar{v}'$ quantifier-free. The $M$ in this notation makes it explicit that the set of events $E$ is taken from the event-B machine $M$. 

\begin{lemma}\label{lem-conv}

Let $M$ be an Event-B machine with state variables $\bar{v} = (v_0, ..., v_m)$ satisfying a convergence property $\mathit{conv}(\varphi)$. Then there exists an Event-B machine $M'$ with state variables $\bar{v}' = (\bar{v}, \bar{w}, l, s, u)$, where $\bar{v}$, $\bar{w} = (w_0, \ldots, w_m)$, $l$, $s$ and $u$ are pairwise different variables, and a term $t(\bar{v}, \bar{w}, l, s, u)$ such that: 

\begin{enumerate}

\item The formulae $\mathit{var}_c(M, t, \varphi)$ is valid for $M'$.  
 
\item For each computation $\sigma$ of $M$ there exists a computation $\sigma'$ of $M'$ with $\sigma = \sigma'|_{s=1,\bar{v}}$ and vice versa, where $\sigma'|_{s=1,\bar{v}}$ is the sequence of states resulting from the selection of those with $s = 1$ and projection to the state variables $\bar{v}$.

\end{enumerate}

\end{lemma}

Before entering into the proof let us briefly sketch the key idea, which is as follows. The machine $M'$ uses a copy $\bar{w}$ of the state variables, and events are first executed on them. Furthermore, $M'$ uses a status variable $s$. As long as the machine $M'$ is in a state satisfying $\varphi$ and $s=0$ holds, the tuple $\bar{w}$ is added to a list $l$. In case of a state satisfying $\neg\varphi$ the machine will switch to status $s=1$. As there can only be finite subsequences of states satisfying $\varphi$, such a status switch will occur, unless the machine runs into a deadlock. When the status $s=1$ is reached, the collected tuples in the list $\ell$ will be one-by-one copied to the original state variables $\bar{v}$, until finally the machine switches back to status $s=0$. In this way all firing of events are first executed on a copy $\bar{w}$, and the length of the list $l$ defines the desired variant term, which is reset by each status switch from $s=0$ to $s=1$. In addition, a few subtle cases concerning deadlocks need to be considered separately, for which another status variable $u$ is used.

\begin{proof}

In the following, if $F$ is a formula or set of actions and $v, w$ are variables, we use $F(v/w)$ to denote the formula or set of actions obtained by substituting in $F$ each occurrence of $v$ by $w$. Furthermore, we use list and list operators, with~$[]$ denoting the empty list and $l_1 \cdot l_2$ denoting concatenation of lists $l_1$ and $l_2$. Of course, with proper encoding we could use sets instead of lists, but lists give us a  cleaner presentation for this proof. 

Let $M'$ be the Event-B machine built form $M$ as follows: 

\begin{itemize}

\item The initial event $\mathit{init}'$ of $M'$ is the result of first applying the substitution $\bar{v}/\bar{w}$ to the initial event $\mathit{init}$ of $M$, and then adding the following actions: $s, u := 0$, $\bar{v} :\!| \, \neg\varphi(\bar{v}')$ and $l := []$.

\item We add to the set of events $E'$ of $M'$ the following event, which is triggered only once and only if the corresponding computation of $M$ is of length $1$. \\ 
    $e_{\mathit{firstA}} = {\bf any} \, \bar{x} \, {\bf where} \, u = 0 \wedge \neg \big( \bigvee_{e_i \in E} G_{i}(\bar{x}, \bar{v}/\bar{w}) \big) \, {\bf then} \, \{ s:= 1, u:= 2, \bar{v}:=\bar{w}\}$.

\item We add to the set of events $E'$ of $M'$ the following event, which is triggered only once and only if the corresponding computation of $M$ is of length $> 1$. \\ 
    $e_{\mathit{firstB}} = {\bf any} \, \bar{x} \, {\bf where} \, u = 0 \wedge \big( \bigvee_{e_i \in E} G_{i}(\bar{x}, \bar{v}/\bar{w}) \big) \, {\bf then} \, \{u:= 1\}$.
    
\item For each $e_i = {\bf any} \, \bar{x} \, {\bf where} \, G_i(\bar{x}, \bar{v}) \, {\bf then} \, A_i(\bar{x}, \bar{v}, \bar{v}') \, {\bf end}$ in the set of events $E$ of $M$, we add the following two corresponding events to $E'$ (the set of events of $M'$).\\ 
    $e_{ia} = {\bf any} \, \bar{x} \, {\bf where} \, G_{ia}(\bar{x}, \bar{v}, \bar{w}, l, s, u) \, {\bf then} \, A_{ia}(\bar{x}, \bar{v}, \bar{w}, l, s, u, \bar{v}', \bar{w}', l', s', u')$\\
    $e_{ib} = {\bf any} \, \bar{x} \, {\bf where} \, G_{ib}(\bar{x}, \bar{v}, \bar{w}, l, s, u) \, {\bf then} \, A_{ib}(\bar{x}, \bar{v}, \bar{w}, l, s, u, \bar{v}', \bar{w}', l', s', u')$\\
    $e_{ic} = {\bf any} \, \bar{x} \, {\bf where} \, G_{ic}(\bar{x}, \bar{v}, \bar{w}, l, s, u) \, {\bf then} \, A_{ic}(\bar{x}, \bar{v}, \bar{w}, l, s, u, \bar{v}', \bar{w}', l', s', u')$\\
    where
    
\begin{itemize}

\item $G_{ia}(\bar{x}, \bar{v}, \bar{w}, l, s, u) \equiv G_i(\bar{x}, \bar{v}/\bar{w}) \wedge \varphi(\bar{v}/\bar{w}) \wedge s = 0 \wedge u = 1$\\
        $A_{ia}(\bar{x}, \bar{v}, \bar{w}, l, s, u, \bar{v}', \bar{w}', l', s', u') =$ \\ 
        \hspace*{2cm} $A_i(\bar{x}, \bar{v}/\bar{w}, \bar{v}'/\bar{w}') \cup \{l := l \cdot [\bar{w}]\}$ 

\item $G_{ib}(\bar{x}, \bar{v}, \bar{w}, l, s, u) \equiv G_i(\bar{x}, \bar{v}/\bar{w}) \wedge \neg \varphi(\bar{v}/\bar{w}) \wedge s = 0 \wedge u = 1 \wedge l \neq []$\\
        $A_{ib}(\bar{x}, \bar{v}, \bar{w}, l, s, u, \bar{v}', \bar{w}', l', s', u' ) = $\\ \hspace*{2cm} $A_i(\bar{x}, \bar{v}/\bar{w}, \bar{v}'/\bar{w}') \cup \{s := 1, \bar{v} := \mathit{head}(l), l := \mathit{tail}(l) \cdot [\bar{w}] \}$

\item $G_{ic}(\bar{x}, \bar{v}, \bar{w}, l, s, u) \equiv G_i(\bar{x}, \bar{v}/\bar{w}) \wedge \neg \varphi(\bar{v}/\bar{w}) \wedge s = 0 \wedge u = 1 \wedge l = []$\\
        $A_{ic}(\bar{x}, \bar{v}, \bar{w}, l, s, u,  \bar{v}', \bar{w}', l', s', u') = $\\ \hspace*{2cm} $A_i(\bar{x}, \bar{v}/\bar{w}, \bar{v}'/\bar{w}') \cup \{s := 1, \bar{v} := \bar{w}\}$
        
\end{itemize}

\item We add the following two additional events $e_r$ and $e_s$ to the set $E'$ of $M'$: 

\begin{itemize}

\item $G_{r}(\bar{x}, \bar{v}, \bar{w}, l, s, u) \equiv l \neq [] \wedge s = 1 \wedge u = 1$\\
        $A_{r}(\bar{x}, \bar{v}, \bar{w}, l, s, \bar{v}', \bar{w}', l', s') = \{\bar{v} := head(l), l := tail(l) \}$ 

\item $G_{s}(\bar{x}, \bar{v}, \bar{w}, l, s, u) \equiv l = [] \wedge s = 1 \wedge u = 1$\\
        $A_{s}(\bar{x}, \bar{v}, \bar{w}, l, s, u, \bar{v}', \bar{w}', l', s', u') =  \{s := 0\}$ 

\end{itemize}

\end{itemize}

Let $t(\bar{v}, \bar{w}, l, s, u)$ be the term $\mathit{len}(l)$, which gives us the length of the list $l$. We show next that properties (1) and (2) of our lemma hold. Let us start with (1), which can be proven by showing that if $M$ is convergent in $\varphi$, $\sigma$ is a computation of $M'$ and  $0 \leq k \leq \ell(\sigma)$, then $\sigma^{(k)} \models \mathit{var}_c(M, \mathit{len}(l), \varphi)$. Let us assume w.l.o.g. that $\sigma = S_0, S_1, \ldots$ is infinite, i.e., $\ell(\sigma) = \omega$, and proceed by induction on $k$. Note that our proof below also holds when $\sigma$ is finite.

Any computation $\sigma$ of $M'$ will have at least $2$ states. This follows from the fact that $\mathit{init}'$ sets the value of $u$ in $S_0$ to $0$, and thus either event $e_\mathit{firstA}$ or $e_\mathit{firstB}$ will be enabled in $S_0$. Therefore, we start the induction with $k=1$. By definition of $\mathit{init}'$, we have that $S_0 \not\models \varphi(\bar{v})$ and thus that $\sigma^{(0)} \models \mathit{var}_c(M, \mathit{len}(l), \varphi)$. 
If state $S_1$ is the result of triggering $e_\mathit{firstA}$, then by construction of $M'$ from $M$ and the fact that $M$ is convergent in $\varphi$, we get that $S_0 \not\models \varphi(\bar{v}/\bar{w})$. We also get by the actions of $e_\mathit{firstA}$ that the value of $\bar{v}$ in $S_1$ is the same as the value of $\bar{w}$ in $S_0$. Then, $S_1 \not\models \varphi(\bar{v})$. 
If state $S_1$ is the result of triggering $e_\mathit{firstB}$ instead, then the value of $\bar{v}$ is not updated and thus it is the same in both $S_0$ and $S_1$. This again gives us that $S_1 \not\models \varphi(\bar{v})$. We can then conclude that  $\sigma^{(1)} \models \mathit{var}_c(M, \mathit{len}(l), \varphi)$.

If $k+1 > 1$, we know by inductive hypothesis that  $\sigma^{(k)} \models \mathit{var}_c(M, \mathit{len}(l), \varphi)$. Thus, either (a) $S_{k} \not\models \varphi(\bar{v})$ or (b) $S_k \models \forall \bar{x} (\neg \bigvee_{e_i \in E} G_{i}(\bar{x}, \bar{v}))$ or (c) $S_{k} \models \mathit{len}(l) \neq 0 \wedge \mathit{len}(l') < \mathit{len}(l)$, where $l'$ is the value of $l$ in state $S_{k+1}$. In order to show that $\sigma^{(k+1)} \models \mathit{var}_c(M, \mathit{len}(l), \varphi)$, we consider the four different cases depending on the values of $s$ in $S_k$ and $S_{k+1}$. 

\begin{itemize}

\item If $s = 0$ in both $S_{k}$ and $S_{k+1}$, then an  event of the form $e_{ia}$ must have been triggered in $S_{k}$, as otherwise $s$ would be $1$ in $S_{k+1}$ or $S_k$ would be the initial state and thus $k+1 = 1$. We know that $\mathit{len}(l') > \mathit{len}(l)$ whenever an event of the form $e_{ia}$ is triggered. Therefore, option (c) above is false and it must be the case by the inductive hypothesis that $S_{k} \not\models \varphi(\bar{v})$ or $S_k \models \forall \bar{x} (\neg \bigvee_{e_i \in E} G_{i}(\bar{x}, \bar{v}))$. Since the action of an event $e_{ia}$ does not update the value of $\bar{v}$, it follows that $S_{k+1} \not\models \varphi(\bar{v})$ or $S_k \models \forall \bar{x} (\neg \bigvee_{e_i \in E} G_{i}(\bar{x}, \bar{v}))$, and thus $\sigma^{(k+1)} \models \mathit{var}_c(M, \mathit{len}(l), \varphi)$.
    
\item If $s = 1$ in $S_{k}$ and $s = 0$ in $S_{k+1}$, then event $e_s$ was triggered in $S_{k}$, as otherwise $s$ would be $1$ in $S_{k+1}$. We know that $\mathit{len}(l') = \mathit{len}(l)$ since $e_s$ does not update the value of $l$. Therefore, option (c) above is false and it must be the case by the inductive hypothesis that $S_{k} \not\models \varphi(\bar{v})$ or $S_k \models \forall \bar{x} (\neg \bigvee_{e_i \in E} G_{i}(\bar{x}, \bar{v}))$. Since $e_{s}$ neither updates the value of $\bar{v}$, it follows that $S_{k+1} \not\models \varphi(\bar{v})$ or $S_k \models \forall \bar{x} (\neg \bigvee_{e_i \in E} G_{i}(\bar{x}, \bar{v}))$, and thus $\sigma^{(k+1)} \models \mathit{var}_c(M, \mathit{len}(l), \varphi)$.

\item If $s = 0$ in $S_{k}$ and $s = 1$ in $S_{k+1}$, then  event $e_\mathit{last}$ or an event of the form $e_{ib}$ or $e_{ic}$ was triggered in $S_{k}$ (notice that $e_\mathit{firstA}$ can only be triggered in state $S_0$). Otherwise $s$ would be $0$ in $S_{k+1}$. 
We know that if event $e_\mathit{last}$ was triggered in $S_k$, then $S_k \models \forall \bar{x} (\neg  (\bigvee_{e_i \in E} G_{i}(\bar{x}, \bar{v}/\bar{w})))$ and that the value of $\bar{v}$ in $S_{k+1}$ equals the value of $\bar{w}$ in $S_k$. Thus, $S_{k+1} \models \forall x (\neg  (\bigvee_{e_i \in E} G_{i}(\bar{x}, \bar{v})))$. Therefore, $\sigma^{(k+1)} \models \mathit{var}_c(M, \mathit{len}(l), \varphi)$. 
If instead an event of the form $e_{ib}$ was triggered in $S_k$, then $l \neq []$ in $S_{k+1}$. Then, the only event that can be triggered in $S_{k+1}$ is $e_r$ and thus $\mathit{len}(l') < \mathit{len}(l)$. Consequently,  $\sigma^{(k+1)} \models \mathit{var}_c(M, \mathit{len}(l), \varphi)$.
Finally, if an event of the form $e_{ic}$ was triggered in $S_k$, then we know that $S_k \not\models \varphi(\bar{v}/\bar{w})$ and that the value of $\bar{v}$ in $S_{k+1}$ is equal to the value $\bar{w}$ in $S_k$. Therefore, $S_{k+1} \not\models \varphi(\bar{v})$, which implies that $\sigma^{(k+1)} \models \mathit{var}_c(M, \mathit{len}(l), \varphi)$. 

\item If $s = 1$ in both $S_{k}$ and $S_{k+1}$, then event $e_r$ was triggered in $S_{k}$, as otherwise $s$ would be $0$ in $S_{k+1}$. If $\mathit{tail}(l) \neq []$ in $S_{k}$, then only event $e_r$ can be triggered in $S_{k+1}$. Triggering event $e_r$ always results in $\mathit{len}(l') < \mathit{len}(l)$. Consequently,  $\sigma^{(k+1)} \models \mathit{var}_c(M, \mathit{len}(l), \varphi)$.
On the other hand, if $\mathit{tail}(l) = []$ in $S_{k}$, then the value of $\bar{v}$ in $S_{k+1}$ equals the value of $\mathit{head}(l)$ in $S_k$. This value corresponds to the value $w$ in the latest preceding state $S_i$ in the computation $\sigma$ where an event of the form $e_{ib}$ was triggered, meaning that $S_i \not\models \varphi(\bar{v}/\bar{w})$. Thus,  $S_{k+1} \not\models \varphi(\bar{v})$ and consequently $\sigma^{(k+1)} \models \mathit{var}_c(M, \mathit{len}(l), \varphi)$. 

\end{itemize}

Next we show that property (2) also holds, i.e., that $\sigma$ is a computation of $M$ iff $\sigma = \sigma'|_{s = 1,\bar{v}}$ for some computation $\sigma'$ of $M'$. We use the fact that $M$ is convergent in $\varphi$ to divide every computation $\sigma$ of $M$ into contiguous segments $s_0, s_1, \ldots$, where for each segment $s_j = S_{j0} \ldots S_{jk_j}$ the following holds:
\begin{itemize}
    \item The first state in the segment $s_{j+1}$ (if it exists) is the immediate successor in $\sigma$ of the last state $S_{jk_j}$ in segment $s_j$.
    \item $S_{jk_j} \not\models \varphi(\bar{v})$.
    \item $S_{jm} \models \varphi(\bar{v})$ for all $0 \leq m < k_j$.
\end{itemize}

It follows that for every segment $s_j$ of a computation $\sigma$ of $M$ there is a corresponding segment $s_j'$ of a computation $\sigma'$ of $M'$ such that $s_j = s_j'|_{s=1,\bar{v}}$, and vice versa. Next we prove the first direction of this fact by induction on $j$. 

For $j = 0$, let the initial segment $s_0 = S_{00} \ldots S_{0k_0}$. By construction of $\mathit{init}'$, we get that in the initial state $S_{00}'$ of $M'$ the value of $\bar{w}$ coincides with that of $\bar{v}$ in $S_{00}$. If $k_0 = 0$, then depending on whether  $\ell(\sigma) = 1$ or $\ell(\sigma) > 1$, either event $e_\mathit{firstA}$ or $e_\mathit{firstB}$ can be triggered in $S_{00}'$. In the former case, $s = 1$ in the next state $S_{01}'$ in the computation of $M'$ and the value of $\bar{v}$ coincides with its value in $S_0$. A last step in the computation of $M'$ triggers $e_s$, which leads to a further state $S_{02}'$ where $s = 0$. Thus, $s_0 = s_0'|_{s = 1, \bar{v}}$ for $s_0' = S_{00}' S_{01}' S_{02}'$. In the latter case, the value of $\bar{w}$ and $s$ in state $S_{01}'$ coincide with the respective values in $S_{00}'$ and  an event of the form $e_{ic}$ will be triggered (notice that $l = []$ and an event $e_i$ was triggered by $M$ in state $S_{00}$ as $\ell(\sigma) > 1$). Thus, in the next state $S'_{02}$ in the computation of $M'$ we get that $s = 1$ and the value of $\bar{v}$ coincides with its value in $S_{00}$. An additional step of $M'$ is triggered in state $S_{02}'$ by the event $e_s$ leading to state $S_{03}'$ where $s = 0$. Thus $s_0 = s_0'|_{s = 1, \bar{v}}$ for $s_0' = S_{00}'S_{01}'S_{02}'S_{03}'$.  

On the other hand, if in the initial segment $s_0$ of $\sigma$ we have that $k_0 > 0$, then we know by construction of $M'$ that in its initial state $S_{00}'$ event $e_\mathit{firstB}$ will be triggered, and that there is a computation of $M'$ in which it holds that the value of $\bar{w}$ in $S_{0m}'$ for each $1 \leq m \leq k_0+1$ coincides with the value of $\bar{v}$ in $S_{0m-1}$. Moreover, we know that an event of the form $e_{ia}$ is then triggered in each of these states except for the last one $S_{0k_0+1}$. Consequently, in state $S_{0k_0+1}'$ the list $l$ is of length $k_0$ and the element in the $m$-th position of $l$ corresponds to the value taken by $\bar{v}$ in state $S_{0m-1}$. Furthermore, the state $S_{0k_0+1}'$ in this computation of $M'$ will necessarily trigger an event of the form $e_{ib}$, resulting in a state $S_{0k_0+2}'$ where $s = 1$ and the value of $\bar{v}$ corresponds to its value in $S_{00}$. Clearly, the values that $\bar{v}$ will take in the next $k_0$ states following $S_{0k_0+2}'$ in this computation of $M'$ will correspond to its values in $S_{01}, \ldots S_{0k_0}$, respectively. In the last of these states, $M'$ will trigger event $e_s$ leading to a further state where $s$ takes again he value $0$. Thus $s_0 = s_0'|_{s = 1, \bar{v}}$, where $s_0'$ is the computation of $M'$ of length $2k_0+3$ described above.

For the inductive step we assume that $j + 1 > 0$. The inductive hypothesis gives us that for the segments  $s_j$ and $s_j'$ of $M$ and $M'$, respectively, it holds that $s_j = s_j'|_{s = 1, \bar{v}}$. It is easy to see that the values of $\bar{w}$ in the last state of $S_{jk_j'}'$ of $s_j'$ coincide with the value of $\bar{v}$ in the last state $S_{jk_j}$ of $s_j$. Since furthermore $s = 0$ in $S_{jk_j'}'$, we can use a similar argument as before to show that $s_{j+1} = s_{j+1}'|_{s = 1, \bar{v}}$. We omit the tedious but simple technical details.

Likewise, we can prove (by induction on $j$) that for every segment $s_j'$ of a computation $\sigma'$ of $M'$ there is a corresponding segment $s_j$ of a computation $\sigma$ of $M$ such that $s_j = s_j'|_{s=1,\bar{v}}$. Again, we omit the tedious, but simple technical details.\qed

\end{proof}

\begin{lemma}\label{lem-div}

Let $M$ be an Event-B machine with state variables $\bar{v} = (v_0, ..., v_m)$ satisfying a divergence property $\mathit{div}(\varphi)$. Then there exists an Event-B machine $M'$ with state variables $\bar{v}' = (\bar{v}, \bar{c}, \bar{b})$, where $\bar{c} = (c_0, \ldots, c_m)$, $\bar{b} = (b_0, \ldots, b_m)$ and $\bar{v}$ are pairwise different variables, and a term $t(\bar{v}, \bar{c}, \bar{b})$ such that: 

\begin{enumerate}

\item The formulae $\mathit{var}_d(M', t, \varphi)$ is valid for $M'$;
 
\item For each computation $\sigma$ of $M$ there exists a computation $\sigma'$ of $M'$ with $\sigma = \sigma'|_{\bar{v}}$ and vice versa, where $\sigma'|_{\bar{v}}$ is the sequence of states resulting from the projection to the state variables $\bar{v}$.

\end{enumerate}

\end{lemma}

Again before entering into the proof let us sketch the key idea as follows. In $M'$ for each state variable $v_i$ we add a state variable $c_i$, which we initialise as $\emptyset$, and into which a new value $v'$ is inserted, whenever the machine makes a step in a state satisfying $\neg\varphi$. Then the tuple $(c_0, \dots, c_m)$ is a monotone increasing sequence of tuples of sets, which are represented in $\text{HF}(A)$ using the common Kuratowsky encoding. As in every trace there are only finitely many states satisfying $\neg\varphi$, there exists a maximum tuple that will never be exceeded. In $M'$ we initialise $(b_0, \dots, b_m)$ by these maximum set values, and never update these state variables. As the order $\le$ on $\text{HF}(A)$ subsumes set inclusion, we can see that $\#(b_0, \dots, b_m) - \#(c_0, \dots, c_m)$ defines the desired variant term.

\begin{proof}

As $M \models \Div(\varphi)$ holds, in any trace $\sigma$ there can only be finitely many states satisfying $\neg\varphi$. Consequently, for each state variable $v_i$ there are only finitely many values that $v_i$ can take in any state satisfying $\neg\varphi$ of any trace. Let $M_i \in \text{HF}(A)$ be the set of all these values.

We define the machine $M'$ with additional state variables $\bar{c} = (c_0, \ldots, c_m)$ and $\bar{b} = (b_0, \ldots, b_m)$. The initialisation event of $M'$ simply extends the event \textit{init} of $M$ by the assigments $c_i := \emptyset$ and $b_i := M_i$. Then the projection of the initial state $S_0$ of $M'$ to the state variables $\bar{v}$ is the initial state of $M$.

For each event $e$ of $M$ with guard $G_e(\bar{x}, \bar{v})$ and action $A_e(\bar{x}, \bar{v}, \bar{v}')$ we define two events $e_+$ and $e_-$ of $M'$. 
The guard of $e_+$ is $G_{e_+}(\bar{x}, \bar{v}, \bar{c}, \bar{b}) = G_e(\bar{x}, \bar{v}) \wedge \varphi(\bar{v})$, and the action $A_{e_+}(\bar{x}, \bar{v}, \bar{c}, \bar{b}, \bar{v}', \bar{c}', \bar{b}')$ is the same as $A_e(\bar{x}, \bar{v}, \bar{v}')$.
The guard of $e_-$ is $G_{e_-}(\bar{x}, \bar{v}, \bar{c}, \bar{b}) = G_e(\bar{x}, \bar{v}) \wedge \neg\varphi(\bar{v})$, and the action $A_{e_-}(\bar{x}, \bar{v}, \bar{c}, \bar{b}, \bar{v}', \bar{c}', \bar{b}')$ results from adding the assignments $c_i := c_i \cup \{ v_i^\prime \}$ for all $i=1,\dots,m$ to the action $A_e(\bar{x}, \bar{v}, \bar{v}')$.

Regardless if $M'$ is in a state satisfying $\varphi$ or in a state satisfying $\neg\varphi$, the projection of the action of either $e_+$ or $e_-$ to the state variables $\bar{v}$ yields the action of $e$. It follows that property (ii) of the lemma holds.

For property (i) we define the variant term $t(\bar{v}, \bar{c}, \bar{b}) = (b_0 - c_0, \dots, b_m - c_m)$. The values of the state variables $\bar{b}$ are never updated, and the values of $\bar{c}$ are only updated by events $e_-$, i.e. only in states satisfying $\neg\varphi$. This implies that
\begin{equation}\label{eq-div3}
\varphi(\bar{v}) \wedge G_e(\bar{x}, \bar{v}, \bar{c}, \bar{b}) \wedge P_{A_e}(\bar{x}, \bar{v}, \bar{c}, \bar{b}, \bar{v}', \bar{c}', \bar{b}') \rightarrow t(\bar{v}', \bar{c}', \bar{b}') = t(\bar{v}, \bar{c}, \bar{b})
\end{equation}
holds for all events $e$ of $M'$ and all $\bar{x}$, $\bar{v}$, $\bar{c}$, $\bar{b}$, $\bar{v}'$, $\bar{c}'$, $\bar{b}'$.

As every event $e_-$ inserts additional values into the sets $c_i$, the sequence of values for the tuple $(c_0, \dots, c_m)$ is strictly increasing in states satisfying $\neg\varphi$. We cannot have equality, because otherwise we would have a state $S_i$ satisfying $\neg\varphi$ with a possible successor state $S_{i+1} = S_i$ resulting from firing some event $e_-$. Then this event could be fired again and again, thus leading to a trace with an infinite subsequence of states satisfying $\neg\varphi$ contradicting $M \models \Div(\varphi)$. Then the sequence of tuples $(b_0 - c_0, \dots, b_m - c_m)$ is strictly decreasing in states satisfying $\neg\varphi$, i.e. 
\begin{equation}\label{eq-div1}
\neg\varphi(\bar{v}) \wedge G_e(\bar{x}, \bar{v}, \bar{c}, \bar{b}) \wedge P_{A_e}(\bar{x}, \bar{v}, \bar{c}, \bar{b}, \bar{v}', \bar{c}', \bar{b}') \rightarrow t(\bar{v}', \bar{c}', \bar{b}') < t(\bar{v}, \bar{c}, \bar{b})
\end{equation}
holds for all events $e$ of $M'$ and all $\bar{x}$, $\bar{v}$, $\bar{c}$, $\bar{b}$, $\bar{v}'$, $\bar{c}'$, $\bar{b}'$. 

Furthermore, the choice of the constants $M_i$ stored in the state variable $b_i$ ensures that always $c_i \subsetneq b_i$ holds, which implies 
\begin{equation}\label{eq-div2}
\neg\varphi(\bar{v}) \wedge G_e(\bar{x}, \bar{v}, \bar{c}, \bar{b}) \wedge P_{A_e}(\bar{x}, \bar{v}, \bar{c}, \bar{b}, \bar{v}', \bar{c}', \bar{b}') \rightarrow t(\bar{v}, \bar{c}, \bar{b}) \neq \emptyset
\end{equation}
for all events $e$ of $M'$ and all $\bar{x}$, $\bar{v}$, $\bar{c}$, $\bar{b}$, $\bar{v}'$, $\bar{c}'$, $\bar{b}'$. 

Finally, (\ref{eq-div1}), (\ref{eq-div2}) and (\ref{eq-div3}) together are equivalent to claim (i), which concludes the proof.\qed

\end{proof}

Lemmata \ref{lem-conv} and \ref{lem-div} imply that adding the derivation rules CONV and DIV to a sound and complete set of derivation rules for first-order logic yields a sound and complete set of derivation rules for the derivation of convergence and divergence formulae, provided that the Event-B machines are conservatively extended to guarantee the existence of the required variants. Of course, for a proof of a convergence or divergence property the appropriate variant must be defined; the lemmata above only guarantee that this is always possible.

Note that the extensions exploited in the lemmata above are indeed $(1,n)$-refinements. We therefore say that an Event-B machine $M$ is {\em sufficiently refined} with respect to a finite set $\Phi \subseteq \mathcal{L}$ of state formulae iff for every $\varphi \in \Phi$ there exist terms $t_c$ and $t_d$ such that the variant formulae $\text{var}_c(t_c,\varphi)$ and $\text{var}_d(t_d,\varphi)$ are valid for $M$. We summarise this result in the following theorem.

\begin{theorem}\label{thm-variants}

Let $\mathfrak{R}$ be a sound and complete set of derivation rules for first-order logic with types. Then $\mathfrak{R} \cup \{ {\rm CONV}, {\rm DIV} \}$ is a sound and complete set of derivation rules for the proof of transition, convergence, divergence and deadlock-freeness formulae $\Leadsto(\varphi_1, \varphi_2)$, $\Conv(\varphi)$, $\Div(\varphi)$ and $\Dlf(\varphi)$ for Event-B machines that are sufficiently refined with respect to $\{ \varphi \}$.

\end{theorem}

\section{The $\square$LTL Fragment}\label{sec:fragment}

We now investigate a fragment of LTL(EB), in which selected liveness conditions can be expressed. Of particular interest are the following conditions:

\begin{description}

\item[Invariance.] An {\em invariance condition} is given by an LTL(EB) formula of the form $\square \, \varphi$ for some state formula $\varphi$. Such a condition is satisfied by a machine $M$ iff all states $S$ in all traces of $M$ satisfy $\varphi$.

\item[Existence.] An {\em existence condition}---we also call it an {\em unconditional progress condition}---is given by an LTL(EB) formula of the form $\square \lozenge \varphi$ for some state formula $\varphi$. Such a condition is satisfied by a machine $M$ iff in every infinite trace of $M$ there is an infinite subsequence of states satisfying $\varphi$, and every finite trace of $M$ terminates in a state satisfying $\varphi$.

\item[Progress.] A {\em progress condition}---to emphasise the difference to existence conditions we also call it a {\em conditional progress condition}---is given by an LTL(EB) formula of the form $\square (\varphi \rightarrow\lozenge \psi)$ for some state formula $\varphi$. Such a condition is satisfied by a machine $M$ iff in every trace of $M$ a state $S_i$ satisfying $\varphi$ is followed some time later by a state $S_j$ ($j \ge i$) satisfying $\psi$. A progress condition with $\varphi = \textbf{true}$ degenerates to an existence condition.

\item[Persistence.] A {\em persistence condition} is given by an LTL(EB) formula of the form $\lozenge \square \, \varphi$ for some state formula $\varphi$. Such a condition is satisfied by a machine $M$ iff every infinite trace of $M$ ends with an infinite sequence of states satisfying $\varphi$, i.e. for some $k$ we have $S_i \models \varphi$ for all $i \ge k$, and every finite trace of $M$ terminates in a state satisfying $\varphi$.

\end{description}

It therefore makes sense to define the sought fragment of LTL(EB) as the closure of the set $\mathcal{L}$ of state formulae under a more restrictive set of constructors, which basically have the form of the formulae above. We denote this fragment as $\square$LTL. The set of well-formed formulae (wff) of $\square$LTL is inductively defined by the following rules:   

\begin{itemize}

\item If $\varphi \in {\cal L}$, then $\varphi$ is a $\square$LTL wff. 

\item If $\varphi$ is a $\square$LTL wff, then $\square \varphi$, $\square \Diamond \varphi$ and $\Diamond \square \varphi$ are $\square$LTL wff.

\item If $\varphi$ is a state formula and $\psi$ is a $\square$LTL wff, then $\square (\varphi \rightarrow \Diamond \psi)$ is a $\square$LTL wff. 

\item Nothing else is a $\square$LTL wff.
    
\end{itemize}

In the remainder of this section we will present a sound set of derivation rules for $\square$LTL, which modify and extend the derivation rules discovered by Hoang and Abrial \cite{hoang:icfem2011}. We want to show that together with derivation rules for first-order logic with types and the rules CONV and DIV from the previous section we obtain a complete set of derivation rules.

However, LTL as a linear time logic is well suited for linear traces, whereas Event-B machines in general are non-deterministic and yield multiple traces with the same start state. This makes it rather unlikely that we could show completeness. Therefore, we restrict the Event-B machines under consideration. 

We say that a machine $M$ is {\em tail-homogeneous} for a state formula $\varphi$ iff $\tilde{M} \models \Conv(\neg\varphi)$ or $\tilde{M} \models \Div(\neg\varphi)$ holds. $M$ is called {\em tail-homogeneous} for set $\Phi$ of state formulae iff it is tail-homogeneous for all $\varphi \in \Phi$.

\subsection{Sound Derivation Rules}

For invariance conditions we can exploit a simple induction principle. In order to establish invariance of a state formula $\varphi$ it suffices to show that $\varphi$ holds in the initial state, and furthermore that for every $1 \leq k < \ell(\sigma)$ whenever $\sigma^{(k)} \models \varphi$ holds, then also $\sigma^{(k+1)} \models \varphi$ holds. In addition, invariance of a state formula $\varphi$ implies invariance of a weaker state formula $\psi$. This gives rise to the following lemma.

\begin{lemma}\label{lem-invariance}

For state formulae $\varphi, \psi \in \mathcal{L}$ the derivation rules\\
\begin{minipage}{6cm}
\begin{prooftree}
\AxiomC{$\vdash \psi_\mathit{init} \rightarrow \varphi$}
\AxiomC{\hspace*{-5mm}$M \vdash \Leadsto(\varphi,\varphi)$}
\LeftLabel{$\mathrm{INV}_1$:}
\BinaryInfC{$M \vdash \square \varphi$}
\end{prooftree}
\end{minipage}
\qquad
\begin{minipage}{5.5cm}
\begin{prooftree}
\AxiomC{$\vdash \varphi \rightarrow \psi$}
\AxiomC{\hspace*{-5mm}$M \vdash \square \varphi$}
\LeftLabel{$\mathrm{INV}_2$:}
\BinaryInfC{$M \vdash \square \psi$}
\end{prooftree}
\end{minipage}
are sound for the derivation of invariance properties in $\square$LTL.

\end{lemma}

For existence conditions we must have that finite traces terminate in a state satisfying $\varphi$, which is the case if $\Dlf(\neg\varphi)$ holds. For infinite traces we must have infinite subsequences of states satisfying $\varphi$, which by definition is the case, if $\Conv(\neg\varphi)$ holds. This establishes the following lemma, which was proven in \cite[p.~462]{hoang:icfem2011}.

\begin{lemma}\label{lem-existence}

For a state formula $\varphi \in \mathcal{L}$ the derivation rule\\
\begin{minipage}{12.5cm}
\begin{prooftree}
\AxiomC{$M \vdash \, \Conv(\neg \varphi)$}
\LeftLabel{$\mathrm{LIVE}$: \quad}
\UnaryInfC{$M \vdash \square \Diamond \varphi$}
\end{prooftree}
\end{minipage}
is sound for the derivation of existence properties in $\square$LTL.

\end{lemma}

For (conditional) progress conditions Hoang and Abrial also showed the soundness of two derivation rules. However, these rules contain arbitrary UNTIL-formulae and thus refer to LTL(EB), but not to the fragment $\square$LTL. Therefore, we need to combine the rules into a single derivation rule, which is further generalised.

\begin{lemma}\label{lem-progress}

For state formulae $\varphi_1, \varphi_2, \varphi_3 \in \mathcal{L}$ the derivation rule\\
\begin{minipage}{12.5cm}
\begin{prooftree}
\AxiomC{$M \vdash \Div(\neg \varphi_3)$}
\AxiomC{\hspace*{-5mm}$M \vdash \Leadsto(\varphi _3 \wedge \neg \varphi_2, \varphi_3 \vee \varphi_2)$}
\AxiomC{\hspace*{-5mm}$M \vdash \square (\varphi_1 \wedge \neg \varphi_2 \rightarrow \varphi_3 )$}
\LeftLabel{$\mathrm{PROG}$:}
\TrinaryInfC{$M \vdash \square (\varphi _1 \rightarrow \Diamond \varphi_2)$}
\end{prooftree}
\end{minipage}
is sound for the derivation of progress properties in $\square$LTL.

\end{lemma}

\begin{proof}

Let $\sigma$ be an arbitrary trace of $M$, and consider $k$ with $\sigma^{(k)} \models \varphi_1$. If also $\sigma^{(k)} \models \varphi_2$ holds, we get $\sigma^{(k)} \models \varphi_1 \rightarrow \lozenge \varphi_2$ by definition.

Otherwise, for $\sigma^{(k)} \models \neg \varphi_2$ the third antecedent of the PROG rule implies $\sigma^{(k)} \models \varphi_3$. The first antecedent of the rule implies that there exists some $m$ such that $\sigma^{(m')} \models \neg \varphi_3$ holds for all $m' \ge m$. 

Let $m$ be minimal with this property, in particular $m > k$. Then the second antecedent of the rule implies that there exists some $\ell < m$ such that $\sigma^{(\ell')} \models \varphi_3 \wedge \neg \varphi_2$ holds for all $k \le \ell' \le \ell$. If we take $\ell$ maximal with this property, we must have that $\sigma^{(\ell)} \models \varphi_2$ holds by the definition of transition formulae. This shows again $\sigma^{(k)} \models \varphi_1 \rightarrow \lozenge \varphi_2$, hence $M \models \square (\varphi_1 \rightarrow \lozenge \varphi_2)$ as claimed.\qed

\end{proof}

For persistence conditions we must have that finite traces terminate in a state satisfying $\varphi$, which is the case if $\Dlf(\neg\varphi)$ holds. Infinite traces must end with an infinite sequence of states satisfying $\varphi$, which by definition is the case, if $\Div(\varphi)$ holds. This establishes the following lemma, which was proven in \cite[p.~464]{hoang:icfem2011}. Note that the antecedent $\Dlf(\neg \varphi)$ can be dropped, if all traces of $M$ are infinite, which is the case for trivial extensions.

\begin{lemma}\label{lem-persistence}

For a state formula $\varphi \in \mathcal{L}$ the derivation rule\\
\begin{minipage}{12.5cm}
\begin{prooftree}
\AxiomC{$M \vdash \, \Div(\varphi)$}
\AxiomC{$M \vdash \, \Dlf(\neg \varphi)$}
\LeftLabel{$\mathrm{PERS}$: \quad}
\BinaryInfC{$M \vdash \Diamond \square \varphi$}
\end{prooftree}
\end{minipage}
is sound for the derivation of persistence properties in $\square$LTL.

\end{lemma}

The sound derivation rules INV$_1$, INV$_2$, LIVE, PROG and PERS in Lemmata \ref{lem-invariance}-\ref{lem-persistence} are restricted to state formulae. In order to be able to derive arbitrary $\square$LTL formulae further derivation rules are needed. The sound derivation rules in the following lemma are well known for LTL, and proofs follow easily from the definitions.

\begin{lemma}\label{lem-aux}

The following derivation rules are sound for the derivation of formulae in $\square$LTL:

\begin{minipage}{6cm}
\begin{prooftree}
\AxiomC{$M \vdash \, \square \varphi$}
\LeftLabel{$\square$: \quad}
\UnaryInfC{$M \vdash \, \square \square \varphi$}
\end{prooftree}
\begin{prooftree}
\AxiomC{$M \vdash \, \square \neg \varphi$}
\LeftLabel{$\square\vee$: \quad}
\RightLabel{for $\varphi \in \mathcal{L}$}
\UnaryInfC{$M \vdash \, \square (\varphi \rightarrow \lozenge \psi)$}
\end{prooftree}
\begin{prooftree}
\AxiomC{$M \vdash \, \square \lozenge \varphi_2$}
\LeftLabel{$\square\lozenge_1$: \quad}
\UnaryInfC{$M \vdash \, \square (\varphi_1 \rightarrow \lozenge \varphi_2)$}
\end{prooftree}
\begin{prooftree}
\AxiomC{$M \vdash \, \lozenge \square \varphi$}
\LeftLabel{$\square\lozenge_2$: \quad}
\UnaryInfC{$M \vdash \, \square \lozenge \square \varphi$}
\end{prooftree}
\begin{prooftree}
\AxiomC{$M \vdash \, \square ( \varphi_1 \rightarrow \lozenge \varphi_2)$}
\LeftLabel{$\square\lozenge_3$: \quad}
\UnaryInfC{$M \vdash \, \square \lozenge \square ( \varphi_1 \rightarrow \lozenge \varphi_2 )$}
\end{prooftree}
\end{minipage}
\qquad
\begin{minipage}{5cm}
\begin{prooftree}
\AxiomC{$M \vdash \, \lozenge \square \varphi$}
\LeftLabel{$\lozenge\square_1$: \quad}
\UnaryInfC{$M \vdash \, \lozenge \square \square \varphi$}
\end{prooftree}
\begin{prooftree}
\AxiomC{$M \vdash \, \square (\varphi_1 \rightarrow \lozenge \varphi_2)$}
\LeftLabel{$\lozenge\square_2$: \quad}
\UnaryInfC{$M \vdash \, \lozenge \square \square (\varphi_1 \rightarrow \lozenge \varphi_2)$}
\end{prooftree}
\begin{prooftree}
\AxiomC{$M \vdash \, \lozenge \square \varphi$}
\LeftLabel{$\lozenge\square_3$: \quad}
\UnaryInfC{$M \vdash \, \lozenge \square \lozenge \square \varphi$}
\end{prooftree}
\begin{prooftree}
\AxiomC{$\tilde{M} \vdash \, \varphi$}
\LeftLabel{{\rm EXT}: \quad}
\UnaryInfC{$M \vdash \, \varphi$}
\end{prooftree}
\end{minipage}

\end{lemma}

The last derivation rule EXT is based on the fact that invariance, existence, progress and persistence conditions hold for a machine $M$ iff they hold for its trivial extension $\tilde{M}$.

\subsection{Completeness}

Taking all our lemmata together we obtain a sound system of derivation rules for $\square$LTL formula, which comprises a sound and complete set of derivation rules $\mathfrak{R}$ for first-order logic with types, the derivation rules CONV, DIV for convergence and divergence, the derivation rules INV$_1$, INV$_2$, LIVE, PROG and PERS for invariance, existence, progress and persistence, and the auxiliary derivation rules for modal formulae in Lemma \ref{lem-aux}. In the following we denote this set of derivation rules as $\mathfrak{R}_{\square\text{LTL}}$. We will now show that $\mathfrak{R}_{\square\text{LTL}}$ is also complete in the sense that $M \models \varphi$ implies $M \vdash_{\mathfrak{R}_{\square\text{LTL}}}$ for machines that are tail-homogeneous with respect to all state formulae appearing in $\varphi$.

\begin{theorem}

The system $\mathfrak{R}_{\square\text{LTL}}$ of derivation rules is sound and complete for $\square$LTL on Event-B machines $M$ that are sufficiently refined and tail-homogeneous with respect to the set $\Phi$ of state formulae appearing as subformulae of the formulae in $\square$LTL to be proven as well as their negations.

\end{theorem}

\begin{proof}

We only need to show the completeness. For this assume that $M \models \varphi$ holds for $\varphi \in \square\text{LTL}$, hance also $\tilde{M} \models \varphi$. We use induction over the nesting depth of modal operators in $\varphi$ to show that $M \vdash_{\mathfrak{R}_{\square\text{LTL}}} \varphi$ holds. In the following let $\sigma$ always be an arbitrary trace of $\tilde{M}$.

For $\varphi \in \mathcal{L}$, transition formulae $\varphi = \Leadsto(\psi,\chi)$, convergence formulae $\varphi = \Conv(\psi)$, divergence formulae $\varphi = \Div(\psi)$ and deadlock-freeness formulae $\varphi = \Dlf(\psi)$ we immediately get $M \vdash_{\mathfrak{R}_{\square\text{LTL}}} \varphi$ from Theorem \ref{thm-variants}. This leaves the remaining cases of invariance formulae $\varphi = \square \psi$, existence formulae $\varphi = \square \lozenge \psi$, progress formulae $\varphi = \square(\psi \rightarrow \lozenge \chi)$, and persistence formulae $\varphi = \lozenge \square \psi$.

\textbf{Invariance.} Consider first an invariance formula $\varphi = \square \psi$ with $\psi \in \mathcal{L}$. Then $\sigma^{(k)} \models \psi$ holds for all $k$. In particular, for $k=0$ we get $\models \psi_{init} \rightarrow \psi$ and hence $\vdash \psi_{init} \rightarrow \psi$, because $\mathfrak{R}$ is complete. Furthermore, by definition $M \models \Leadsto(\psi,\psi)$ holds, and the completeness of $\mathfrak{R}$ implies $M \vdash \Leadsto(\psi,\psi)$. Together with the derivation rule INV$_1$ we get $M \vdash \square \psi$ as claimed.

Next let $\varphi = \square \psi$ with $\psi \notin \mathcal{L}$. 
(1) For $\psi = \square \chi$ we get $\sigma^{(k)} \models \square \chi$ for all $k$ and hence also $\sigma^{(k)} \models \chi$, i.e. $M \models \square \chi$. By induction we get $M \vdash \square \chi$, and the derivation rule $\square$ implies $M \vdash \square \square \chi$ as claimed. 
(2) For $\psi = \square \lozenge \chi$ we get $\sigma^{(k)} \models \square \lozenge \chi$ for all $k$, hence also $\sigma^{(k)} \models \lozenge \chi$ for all $k$ and $M \models \square \lozenge \chi$. By induction we get $M \vdash \square \lozenge \chi$, and the derivation rule $\square$ implies $M \vdash \square \square \lozenge \chi$ as claimed. 
(3) For $\psi = \square (\chi_1 \rightarrow \lozenge \chi_2)$ we get $\sigma^{(k)} \models \square (\chi_1 \rightarrow \lozenge \chi_2)$ for all $k$, hence also $\sigma^{(k)} \models \chi_1 \rightarrow \lozenge \chi_2$ for all $k$, i.e. $M \models \square (\chi_1 \rightarrow \lozenge \chi_2)$. By induction we get $M \vdash \square (\chi_1 \rightarrow \lozenge \chi_2)$, so the derivation rule $\square$ implies $M \vdash \square \psi$ as claimed.
(4) For $\psi = \lozenge \square \chi$ we get $\sigma^{(k)} \models \lozenge \square \chi$ for all $k$, hence also $M \models \lozenge \square \chi$. By induction we get $M \vdash \lozenge \square \chi$, and the derivation rule $\square\lozenge_2$ implies $M \vdash \square \lozenge \square \chi$ as claimed.

\textbf{Existence.} Next consider an existence formula $\varphi = \square \lozenge \psi$ with $\psi \in \mathcal{L}$. Then $\sigma^{(k)} \models \lozenge \psi$ holds for all $k$. If $\sigma$ is infinite, this implies $M \models \Conv(\neg\psi)$. If $\sigma$ is finite, it cannot terminate in a state satisfying $\neg \psi$, hence $M \models \Dlf(\neg\psi)$. Using Theorem \ref{thm-variants} we get $M \vdash \Dlf(\neg\psi)$ and $M \vdash \Conv(\neg\psi)$. Then with the derivation rule LIVE we obtain $M \vdash \varphi$.

Now let $\varphi = \square \lozenge \psi$ be a general existence formula with $\psi \notin \mathcal{L}$. 
(1) For $\psi = \square \chi$ we get $\sigma^{(k)} \models \lozenge \square \chi$ for all $k$, hence also $M \models \lozenge \square \chi$. By induction we get $M \vdash \lozenge \square \chi$, and by using the derivation rule $\square$ this implies $M \vdash \square\lozenge \square \chi$, i.e. $M \vdash \varphi$.
(2) For $\psi = \square \lozenge \chi$ we get $\sigma^{(k)} \models \lozenge \square \lozenge \chi$ for all $k$, hence also $\sigma^{(k)} \models \square \lozenge \chi$ for all $k$, which further implies $M \models \square \lozenge \chi$. By induction we get $M \vdash \square \lozenge \chi$, and with the derivation rule $\square\lozenge_3$ (with $\varphi_1 = \textbf{true}$) we obtain $M \vdash \square \lozenge \square \lozenge \chi$, i.e. $M \vdash \varphi$ as claimed.
(3) For $\psi = \square (\chi_1 \rightarrow \lozenge \chi_2)$ we get $\sigma^{(k)} \models \lozenge \square (\chi_1 \rightarrow \lozenge \chi_2)$ for all $k$, hence also $\sigma^{(k)} \models \square (\chi_1 \rightarrow \lozenge \chi_2)$ for all $k$, which further implies $M \models \square (\chi_1 \rightarrow \lozenge \chi_2)$. By induction we get $M \vdash \square (\chi_1 \rightarrow \lozenge \chi_2)$, and with the derivation rule $\square\lozenge_3$ we obtain $M \vdash \square \lozenge \square (\chi_1 \rightarrow \lozenge \chi_2)$, i.e. $M \vdash \varphi$ as claimed.
(4) For $\psi = \lozenge \square \chi$ we get $\sigma^{(k)} \models \lozenge \lozenge \square \chi$ for all $k$, hence also $\sigma^{(k)} \models \lozenge \square \chi$ for all $k$, i.e. $M \models \square \lozenge \square \chi$ and further $M \models \lozenge \square \chi$. By induction we get $M \vdash \lozenge \square \chi$, and then the derivation rule $\square\lozenge_2$ implies $M \vdash \varphi$ as claimed.

\textbf{Progress.} Next consider a progress formula $\varphi = \square (\varphi_1 \rightarrow \lozenge \varphi_2)$ with $\varphi_1, \varphi_2 \in \mathcal{L}$. Then for every  trace $\sigma$ of $M$ and every $k$ with $\sigma^{(k)} \models \varphi_1$ there exists some $\ell \ge k$ with $\sigma^{(\ell)} \models \varphi_2$. As $M$ is tail-homogeneous with respect to $\varphi_2$, we can distinguish two cases.

\textbf{Case 1.} Assume $\tilde{M} \models \Conv(\neg\varphi_2)$ holds. Then $\sigma \models \square \lozenge \varphi_2$ holds for all traces of $\tilde{M}$. Hence $M \models \square \lozenge \varphi_2$ and by induction $\tilde{M} \vdash \square \lozenge \varphi_2$. Then we can apply rule $\square\lozenge_1$ to obtain $M \vdash \square (\varphi_1 \rightarrow \lozenge \varphi_2)$.

\textbf{Case 2.} Assume that $\tilde{M} \models \Div(\neg\varphi_2)$ holds. Then all traces of $\tilde{M}$ contain only a finite number of states satisfying $\varphi_2$, thus we consider all states in these traces that satisfy $\neg \varphi_2$ and for which there exists a later state satisfying $\varphi_2$. Let $\varphi_3$ be a common invariant for these states. More precisely, for a state $S$ satisfying $\neg \varphi_2$ take its type, i.e. the set of all state formulae that hold in $S$, and let $\varphi_S$ be an isolating formula for this type. Then define $\varphi_3$ as the disjunction of all these formulae $\varphi_S$. Then by construction $M \models \square (\varphi_1 \wedge \neg \varphi_2 \rightarrow \varphi_3)$ holds.

As an infinite tail of any trace contains only states satisfying $\neg \varphi_2$, these states must also satisfy $\neg \varphi_1$. Then our construction yields $M \models \square (\varphi_1 \wedge \neg \varphi_2 \rightarrow \varphi_3)$.

Furthermore, as $M$ is sufficiently refined and $\tilde{M}$ is divergent in $\neg \varphi_2$, by Lemma \ref{lem-div} we get a variant term $t(\bar{v})$ such that $\text{var}_d(t, \neg \varphi_1 \wedge \neg \varphi_2)$ is valid, in particular, $t(\bar{v}) > v_{\text{min}}$ is included in $\varphi_3$, and $t(\bar{v}) = v_{\text{min}}$ holds in all tailing states satisfying $\neg \varphi_1 \wedge \neg \varphi_2$. This implies $M \models \Div(\neg \varphi_3)$.

By induction we get $M \vdash \square (\varphi_1 \wedge \neg \varphi_2 \rightarrow \varphi_3)$, $M \vdash \Leadsto(\varphi_3 \wedge \neg \varphi_2, \varphi_3 \vee \varphi_2)$ and $M \vdash \Div(\neg \varphi_3)$. Thus, we can apply the rule PROG, which yields $M \vdash \square (\varphi_1 \rightarrow \lozenge \varphi_2)$.

Next consider a progress formula $\varphi = \square (\varphi_1 \rightarrow \lozenge \varphi_2)$ with $\varphi_2 \notin \mathcal{L}$. In case $\tilde{M} \models \Conv(\neg\varphi_1)$ holds, we immediately get $M \models \square\lozenge \varphi_1$ and hence also $M \models \lozenge \varphi_1$. In this case we consider again four cases for $\varphi_2$.
(1) For $\varphi_2 = \square \chi$ it follows that $M \models \lozenge \square \chi$ holds, hence by induction $M \vdash \lozenge \square \chi$. With rule $\square\lozenge_2$ we infer $M \vdash \square \lozenge \square \chi$, and with rule $\square\lozenge_1$ we get $M \vdash \square (\varphi_1 \rightarrow \lozenge \square \chi)$.
(2) For $\varphi_2 = \square\lozenge \chi$ we get $M \models \lozenge \square \lozenge \chi$ and further $M \models \square \lozenge \square \lozenge \chi$ due to the soundness of $\square\lozenge_2$, which holds not only for $\square$LTL. By induction it follows that $M \vdash \square \lozenge \square \lozenge \chi$, and then the application of rule $\square\lozenge_1$ yields $M \vdash \square (\varphi_1 \rightarrow \lozenge \square \lozenge \chi)$.
(3) For $\varphi_2 = \square (\chi_1 \rightarrow \lozenge \chi_2)$ we have $M \models \lozenge \square (\chi_1 \rightarrow \lozenge \chi_2)$, hence also $M \models \square \lozenge \square (\chi_1 \rightarrow \lozenge \chi_2)$. By induction we get $M \vdash \square \lozenge \square (\chi_1 \rightarrow \lozenge \chi_2)$, hence the application of rule $\square\lozenge_1$ yields again $M \vdash \square (\varphi_1 \rightarrow \lozenge \square (\chi_1 \rightarrow \lozenge \chi_2))$.
(4) Finally, for $\varphi_2 = \lozenge\square \chi$ we get $M \models \lozenge \lozenge \square \chi$ and further $M \models \lozenge \square \chi$. The soundness of rule $\square\lozenge_2$ implies $M \models \square\lozenge \square \chi$, hence also $M \models \square\lozenge\lozenge \square \chi$. By induction we conclude $M \vdash \square\lozenge\lozenge \square \chi$, and finally the application of rule $\square\lozenge_1$ yields $M \vdash \square (\varphi_1 \rightarrow \lozenge\lozenge \square \chi)$, i.e. $M \vdash \square (\varphi_1 \rightarrow \lozenge \varphi_2)$.

To conclude the proof for progress formulae we look at the case that $\tilde{M} \models \Div(\neg\varphi_1)$ holds. Then for all traces $\sigma$ there exists some $k$ with $\sigma^{(\ell)} \models \neg\varphi_1$ for all $\ell \ge k$. If no trace $\sigma$ exists with $\sigma^{(m)} \models \varphi_1$ for some $m$, we have $\tilde{M} \models \square\neg\varphi_1$ and hence also $M \models \square\neg\varphi_1$. By induction we obtain $M \vdash \square\neg\varphi_1$, so the application of rule $\square\vee$ implies again $M \vdash \square (\varphi_1 \rightarrow \lozenge \varphi_2)$.

Therefore, we can assume that there exists a trace $\sigma$ with $\sigma^{(m)} \models \varphi_1$. By exploiting equivalences among $\square$LTL formulae $\varphi_2$ can be rewritten to take one of the forms $\square\psi$, $\square\lozenge\psi$, $\lozenge\square\psi$ with either $\psi \in \mathcal{L}$ or $\psi = \square (\chi_1 \rightarrow \lozenge \chi_2)$ with $\chi_1 \in \mathcal{L}$. 

(1) In case $\varphi_2 = \square\psi$ with $\psi \in \mathcal{L}$ there exists a trace $\sigma$ and some $k \ge m$ with $\sigma^{(\ell)} \models \neg\varphi_1$ for all $\ell \ge k$. That is, $\tilde{M} \models \Div(\psi)$, which implies $\tilde{M} \models \lozenge\square \psi$ and $\tilde{M} \models \square\lozenge\square \psi$. By induction we get $\tilde{M} \vdash \square\lozenge\square \psi$, so the application of rule $\square\lozenge_1$ implies $\tilde{M} \vdash \square (\varphi_1 \rightarrow \lozenge\square \psi)$, i.e. $\tilde{M} \vdash \square (\varphi_1 \rightarrow \lozenge \varphi_2)$. An application of rule EXT implies $M \vdash \square (\varphi_1 \rightarrow \lozenge \varphi_2)$. 

(2) In case $\varphi_2 = \square\lozenge\psi$ with $\psi \in \mathcal{L}$ there exists a trace $\sigma$ such that for all $k \ge m$ there exist some $\ell \ge k$ with $\sigma^{(\ell)} \models \psi$. That is, $\tilde{M} \models \Conv(\neg\psi)$ holds, which implies $\tilde{M} \models \square\lozenge\psi$ and hence also $\tilde{M} \models \square\lozenge\varphi_2$. By induction we get $\tilde{M} \vdash \square\lozenge\varphi_2$, and further $\tilde{M} \vdash \square (\varphi_1 \rightarrow \lozenge \varphi_2)$ by applying rule $\square\lozenge_1$. A final application of rule EXT implies $M \vdash \square (\varphi_1 \rightarrow \lozenge \varphi_2)$. 

(3) In case $\varphi_2 = \lozenge\square\psi$ with $\psi \in \mathcal{L}$ we get $\tilde{M} \models \Div(\psi)$ as in (1). Using the same arguments it follows that $\tilde{M} \models \square\lozenge \varphi_2$ holds. By induction we get $\tilde{M} \vdash \square\lozenge \varphi_2$, and then $M \vdash \square (\varphi_1 \rightarrow \lozenge \varphi_2)$ follows from the application of the rules $\square\lozenge_1$ and EXT.

(4) Next consider the case $\varphi_2 = \square (\psi \rightarrow \lozenge \chi)$ with $\psi \in \mathcal{L}$. If $\tilde{M} \models \Conv(\neg\psi)$ holds, then every trace contains an infinite subsequence of states satisfying $\psi$ and hence also an infinite subsequence of states satisfying $\chi$. If $\tilde{M} \models \Div(\neg\psi)$ holds, then for every trace $\sigma$ there exists some $k$ with $\sigma^{(\ell)} \models \neg\psi$ for all $\ell \ge k$. For both cases we can imply $\tilde{M} \models \square\lozenge\square (\psi \rightarrow \lozenge \chi)$, i.e. $\tilde{M} \models \square\lozenge\varphi_2$. By induction we get $\tilde{M} \vdash \square\lozenge\varphi_2$, and $M \vdash \square (\varphi_1 \rightarrow \lozenge \varphi_2)$ follows from the application of the rules $\square\lozenge_1$ and EXT. The cases $\varphi_2 = \square\lozenge (\psi \rightarrow \lozenge \chi)$ and $\varphi_2 = \lozenge\square (\psi \rightarrow \lozenge \chi)$ with $\psi \in \mathcal{L}$ are handled in a completely analogous way.

\textbf{Persistence.} Finally, consider a persistence formula $\varphi = \lozenge\square\psi$, and first assume $\psi \in \mathcal{L}$. For an arbitrary trace $\sigma$, as $M \models \lozenge\square\psi$ holds, it follows that $\sigma$ cannot end in a state satisfying $\neg\psi$ for finite $\sigma$, and for every infinite $\sigma$ there must exist some $k$ with $\sigma^{(\ell)} \models \varphi$ for all $\ell > k$. That is, both $M \models \Div(\psi)$ and $M \models \Dlf(\neg\psi)$ hold. By induction we obtain $M \vdash \Div(\psi)$ and $M \vdash \Dlf(\neg\psi)$, which allows the rule PERS to be applied to yield $M \vdash \lozenge\square\psi$, i.e. $M \vdash \varphi$.

Next consider a persistence formula $\varphi = \lozenge\square\psi$ with $\psi \notin \mathcal{L}$.
(1) For $\psi = \square \chi$ we get $M \models \lozenge \square \square \chi$, hence also $M \models \lozenge \square \chi$. By induction we have $M \vdash \lozenge \square \chi$, so applying rule $\lozenge\square_1$ implies $M \vdash \lozenge \square \square \chi$, i.e. $M \vdash \varphi$.
(2) For $\psi = \square \lozenge \chi$ we get $M \models \lozenge \square \lozenge \chi$ and hence $M \models \square \lozenge \chi$. Taking $\alpha = \textbf{true}$ it follows that $M \models \square (\alpha \rightarrow \lozenge \chi)$ holds. By induction we get $M \vdash \square (\alpha \rightarrow \lozenge \chi)$, which allows rule $\lozenge\square_2$ to be applied yielding $M \vdash \lozenge \square \square \lozenge \chi$, i.e. $M \vdash \varphi$.
(3) For $\psi = \square (\chi_1 \rightarrow \lozenge \chi_2)$ we get $M \models \lozenge \square (\chi_1 \rightarrow \lozenge \chi_2)$ and hence $M \models \square (\chi_1 \rightarrow \lozenge \chi_2)$. By induction we get $M \vdash \square (\chi_1 \rightarrow \lozenge \chi_2)$, and applying rule $\lozenge\square_2$ implies $M \vdash \varphi$.
(4) For $\psi = \lozenge \square \chi$ we get $M \models \lozenge \square \lozenge \square \chi$ and hence $M \models \lozenge \square \chi$. By induction we have $M \vdash \lozenge \square \chi$, which allows rule $\lozenge\square_3$ to be applied. This implies $M \vdash \lozenge \square \lozenge \square \chi$, i.e. $M \vdash \varphi$.\qed

\end{proof}

Note that the requirement that machines should be tail-homogeneous was only used in the proof for progress formulae. We therefore obtain the following.

\begin{corollary}

The system $\mathfrak{R}_{\square\text{LTL}}$ of derivation rules is sound and complete for proofs of invariance, existence and persistence conditions on Event-B machines that are sufficiently refined with respect to the set $\Phi$ of state formulae appearing as subformulae of the formulae in $\square$LTL to be proven.

\end{corollary}

\section{Conclusion}\label{sec:schluss}

In this paper we defined a fragment of the logic LTL(EB) integrating the UNTIL-fragment of LTL and the logic of Event-B, and proved its completeness for Event-B machines that are sufficiently refined and tail-homogeneous. The former restriction guarantees that variant terms required in the proofs exist. It is no loss of generality, as the necessary refinements are an intrinsic part of the Event-B method. The latter restriction tames the non-determinism in Event-B, which enables a smoother integration with LTL. The $\square$LTL fragment supports the mechanical verification of selected liveness conditions such as conditional and unconditional progress or persistence for Event-B. Our development of the theory is generic enough to allow the conclusion that a transfer of the results to other rigorous state-based methods such as TLA$^+$ or ASMs is straightforward.

However, LTL is a linear-time temporal logic, and as such it is appropriate for deterministic systems, but less so for non-deterministic ones. In Event-B there are two sources of non-determinism. First, events may involve the selection of arbitrary values satisfying the before-after predicate. In most cases this form of non-determinism is considered to be linked to an external selection (see the discussion in \cite{boerger:2024}), i.e. the values are merely parameters in the computation. Consequently, in most cases the use of LTL(EB) will suffice. Second, in every state several events may be enabled, and only one is selected to fire. Though the Event-B refinement process emphasises the reduction of this form of non-determinism, it is not well supported by LTL. In particular, conditions for successor states can only be expressed for all successor states. So it seems a good idea to also consider CTL instead of LTL, though it is well known that persistence formulae cannot be expressed using CTL. Nonetheless, an open question for future research is the definition of a complete fragment of CTL(EB).

For ASMs the logic developed by St\"ark and Nanchen is explicitly tailored to deterministic ASMs \cite{staerk:jucs2001}. Thus, a combination with a fragment of LTL and a straightforward carry-over of the $\square$LTL fragment to ASMs is appropriate. For non-deterministic ASMs the problems of a generalisation have been intensively discussed in \cite{boerger:2003}, which also shows that a combination with CTL would be too weak. The logic of non-deterministic ASMs is second-order, but complete with respect to Henkin semantics. This raises the open research question how to define a complete temporal extension, in which selected liveness properties can be expressed.

\bibliographystyle{splncs04}
\bibliography{ltl(eb)}

\end{document}